\documentclass[11pt]{amsart}
\usepackage{latexsym,amsmath,amsthm,amsfonts}
\usepackage[psamsfonts]{amssymb}

\newtheorem{theorem}{\sc Theorem}

\newtheorem{lemma}[theorem]{\sc Lemma}

\newtheorem{property}[theorem]{\sc Property}
\theoremstyle{definition}

\theoremstyle{remark}

\newcommand{\E}{{\rm I}\kern-0.18em{\rm E}}
\newcommand{\R}{{\rm I}\kern-0.18em{\rm R}}

\begin{document}
\baselineskip=18pt

\title[On Walsh code assignment]
{On Walsh code assignment} 
\author[Tsybakov B.S.]{B. S. Tsybakov}
\author[Tsybakov A.B.]{A. B. Tsybakov}
\address{B. S. Tsybakov} 
\email{\rm
boristsybakov@yahoo.com}
\address{A. B. Tsybakov\\ Laboratoire de Statistique, CREST-ENSAE,
3, av. P.Larousse, 92240 Malakoff, France.} 
\email{\rm
alexandre.tsybakov@ensae.fr}


\subjclass{Primary , Secondary }


\date{September 1,  2012}

\maketitle

\bibliographystyle{amsplain}

\begin{abstract}The paper considers the problem of orthogonal variable spreading Walsh-code assignments. The aim of the paper is to provide assignments that can avoid both complicated signaling from the BS to the users and blind rate and code detection amongst a great number of possible codes. The assignments considered here use a partition of all users into several pools. Each pool can use its own codes that are different for different pools. Each user has only a few codes assigned to it within the pool.  We state the problem as a combinatorial one expressed in terms of a binary $n\times k$ matrix {\bf M} where is the number $n$ of users, and $k$ is the number of Walsh codes in the pool. A solution to the problem is given as a construction of~{\bf M}, which has the assignment property defined in the paper. Two constructions of such {\bf M} are presented under different conditions on~$n$ and~$k$. The first construction is optimal in the sense that it gives the minimal number of Walsh codes~$l$ assigned to each user for given~$n$ and~$k$. The optimality follows from a proved necessary condition for the existence of {\bf M} with the assignment property. In addition, we propose a simple algorithm of optimal assignment for the first construction. 
\end{abstract}


\section{Introduction}

A direct-sequence code division multiple-access (CDMA) third generation wireless network (see [1]) employs the orthogonal variable spreading factor (OVSF) Walsh codes [2-7]. In OVSF systems, the mobile stations (MSÕs or users) that require higher transmission rate in the current frame of the forward channel (from the base station (BS) to the MS) should use shorter length codes. Information about which code the BS will use can be signaled to the MS on a dedicated channel or the MS can perform blind rate and code detection. However, the signaling takes extra resources whereas performing the blind rate and code detection is complicated if there is a large number of possible codes out of which the BS has to choose a code for transmission. It is possible to reduce these difficulties by making the number of codes that can be used for transmission to each MS as small as possible (the Walsh codes that can be used for transmission to a particular MS are called below the codes {\it assigned} to this MS). This can be achieved if all users are partitioned into several pools, each pool can use its own codes taken from the set of all available codes, and a user of a pool monitors the pool codes assigned to it only. In what follows, the assigned code units will be called the full rate codes.

Let  $n$  denote the number of mobile stations (users) in a pool,  $k$ denote the total number of available different full rate Walsh codes in the pool, which equals the total number of pool users that can receive voice simultaneously in the same frame interval at the full rate, and  $l$  denote the number of full rate codes available to each pool user. The  $n$  users are numbered by $1,\dots,n$ . The  $k$  available Walsh codes are numbered by $1,\dots,k$. The Walsh code numbers assigned to user   $i$ are denoted by  $b_{ij}\in \{1,\dots,k\}$ ,  $i=1,\dots,n$, $j=1,\dots,l$. For example, if  $l=3$  and  $k\ge 5$ , it can be that   $b_{11}=1$,    $b_{12}=3$, $b_{13}=5$. This means that the Walsh codes 1, 3, and 5 are assigned to user 1. 
The matrix $S=(b_{ij})$,  $i=1,\dots,n$, $j=1,\dots,l$,  exhibiting all the Walsh codes assigned  to all the users is called the {\it assignment table}. 

A given assignment table has the {\it assignment property} if for any different  $i_1,\dots,i_k$  out of  $\{1,\dots,n\}$ there exist $j_1,\dots,j_k$  such that all  
$b_{i_1j_1},\dots,b_{i_kj_k}$ are different. This means that if an assignment table has the assignment property, the BS can choose the different full codes for any  $k$  (or less) MSÕs and simultaneously transmit to these  $k$  (or less) MSÕs in the frame.

Equivalently, we can describe the code assignment to user $i$ by a binary row of length $k$ with $l$ ones and $k-l$ zeros. The $j$th entry of this row is 1 if the Walsh code $j$  is assigned to user $i$, and the $j$th entry is 0 otherwise. For example, if Walsh codes 1, 3, and 5 are assigned to user $i$  then, assuming  $k=5$, the $i$th binary row is $(1\ 0 \ 1\ 0\ 1)$. Denote by  {\bf M} the $n\times k$ matrix composed of such binary rows for users $i=1,\dots,n$. We will say that 
{\bf M} has the assignment property if the corresponding matrix $S$  has the assignment property. 

The following necessary and sufficient condition for the assignment property is straightforward. \vspace{-1.5mm}
\begin{property}\label{0}
{\it The assignment table $S$ and the corresponding binary matrix  {\bf M} have the assignment property if and only if by permutation of any $k$ rows of {\bf M} we can obtain a $k\times k$ matrix with all diagonal entries equal to 1.}
\end{property}

Note that if {\bf M}  is a matrix having the assignment property, then any matrix  obtained by a row and/or column permutation of {\bf M} has the assignment property as well.

In this paper, we study the problem of constructing matrices with the assignment property for given  $n$,  $k$, and $l$. We will embed it in a more general problem by considering any binary $n\times k$ matrices {\bf M} and not necessarily matrices with exactly $l$ entries 1 in each row. The assignment property for such matrices {\bf M} is defined by the condition stated in Property~\ref{0}. 

This paper is organized as follows. In Section~2, we discuss  the connection of the assignment property to Hall's theorem, which will be used in the proofs of the main results.  In Section~3, we introduce two matrices {\bf M} and, with the help of Hall's theorem, we prove that they satisfy the assignment property.  The first matrix, that we call the {\it $l$-banded matrix}, satisfies the assignment property for odd $k\ge3$ and $k\le n\le 2k$ while the second one, called the {\it augmented $l$-banded matrix}, satisfies this property for even $k\ge4$ and $k\le n\le 2(k-1)$. In Section~4, we propose an algorithm of assignment for the first matrix. This also gives a constructive proof of the assignment property independent of Hall's theorem. In Section~5 we show that the first matrix is optimal and the second one is asymptotically optimal (as $k\to \infty$) in the sense that they have the minimal number of entries 1 for given $n$ and $k$.


\section{Connection to Hall's theorem}

First, recall the following theorem due to Hall (see, e.g., \cite{hall}). Let $S_1,\dots, S_k$ be $k$ subsets of some set~$S$.
We say that different representatives exist in $\{S_i\}$ if one can extract from each $S_i$ one element such that all the $k$ extracted elements are different.
\begin{theorem}\label{th_hall} {\sc (Hall's theorem)} Let $S_1,\dots, S_k$ be $k$ subsets of some set $S$.
Then different representatives exist in $\{S_i\}$ if and only if for any $m\in\{1,\dots, k\}$ and any sequence of different indices $i_1,\dots, i_m$ the union of sets $S_{i_1}\cup\cdots\cup S_{i_m}$ contains at least $m$ different elements.
\end{theorem}
We now apply Theorem~\ref{th_hall} in our setting. Then the set $S$ is the assignment table: $S=\{b_{ij}, i=1,\dots,n, \ j=1\dots, l\}$, where the elements $b_{ij}$ take values in $\{1,\dots, k\}$. The subsets $S_i$ are rows of the assignment table; thus each $S_i$ is a set of $l$ elements $S_i=\{b_{ij}, \ j=1\dots, l\}$. With this notation, the {\it assignment property} is equivalent to the existence of different representatives in $\{S_i\}$. Therefore, to verify whether the assignment property holds for some given table $S$, it is enough to check the condition of Hall's theorem. 
This condition can be restated in terms of binary matrix ${\bf M}$ as shown  in the next lemma. First, introduce some notation. Let ${\bf M}_{i}$ denote the $i$th row of matrix ${\bf M}$. A column of any matrix will be called null column if all its elements are zero, and will be called non-null column otherwise.
\begin{lemma}\label{lem1} The table $S$ (equivalently, the corresponding matrix ${\bf M}$) has the assignment property if and only if, for any $m\in\{1,\dots, k\}$ and any sequence of different indices $i_1,\dots, i_m$,  the number of non-null columns of matrix
$$
{\bf A}(i_1,\dots, i_m)\triangleq\left(\begin{array}{c} 
{\bf M}_{i_1}\\ \cdot \\ \cdot \\ \cdot \\
{\bf M}_{i_m}\\ \end{array}\right)
$$
is not less than $m$.
\end{lemma}
{\sc Proof.} Matrix ${\bf M}$ is an $n \times k$ matrix with  elements
$$
M_{it}={\bf I}\big(\exists \ j\in\{1,\dots, l\}: \ b_{ij}=t\big), \qquad i=1,\dots,n, \ t=1,\dots, k,
$$
where ${\bf I}(\cdot)$ denotes the indicator function and $b_{ij}$ are elements of the assignment table $S$. Note that 
$$
S_{i_1}\cup\cdots\cup S_{i_m}=\big\{b_{ij}: j\in\{1,\dots, l\}, i\in \{i_1,\dots, i_m\}\big\}.
$$
The $t$th column of matrix ${\bf A}(i_1,\dots, i_m)$ is non-null if and only if we have:
$$
\exists \, i\in \{i_1,\dots, i_m\} \ \text{such that} \ M_{it}=1.
$$
In turn, this is equivalent to the condition
$$
\exists \ j\in\{1,\dots, l\}, \ \exists \, i\in \{i_1,\dots, i_m\} \ \text{such that} \ b_{ij}=t.
$$
Therefore, the number of different non-null columns of ${\bf A}(i_1,\dots, i_m)$ is equal to the number of different elements $b_{ij}$ in the set
$S_{i_1}\cup\cdots\cup S_{i_m}$. Thus, the desired result follows from Theorem~\ref{th_hall} and the remarks after it. 


\section{Two constructions of matrix ${\bf M}$}

In this section, we propose two $n\times k$ matrices ${\bf M}$ satisfying the assignment property. The first matrix, that we call the $l$-banded matrix, is constructed for odd values of $k$ while the second one, called the augmented $l$-banded matrix, is constructed for even $k$. At the beginning, we define them with fixed $n$ depending on $k$ (maximal $n$ for each of the two constructions) but the results extend in an obvious way to some range of smaller $n$, cf. the remark at the end of this section.

\medskip

{\bf First construction: the $l$-banded matrix {\bf M}} 

Let $k\ge3$ be an odd number and $n=2k$. Set $l=(k+1)/2$. The binary  $n\times k$ matrix  {\bf M} is called the  $l$-banded matrix if its $j$th row  is the  $(j-1)$  position rightward cyclic shift of row $(\underbrace{1\dots1}_l \underbrace{\ 0\dots 0}_{k-l})$ whose first $l$ entries are equal to 1  and the remaining $k-l=(k-1)/2$ entries are 0. An example of such matrix {\bf M} is given in Figure 1 (left). 

\medskip

{\bf Second construction: the augmented $l$-banded matrix {\bf M}}. 

Let $k\ge 4$ be an even number and $n=2(k-1)$. Set $l=k/2$. A binary  $n\times k$ matrix  {\bf M}, called the augmented $l$-banded matrix, is defined 
as follows: ${\bf M}=[{\bf M}' \ {\bf b}]$ where ${\bf M}'$ is the $l$-banded matrix with $l=k/2$ (${\bf M}'$ is a $n\times (k-1)$ matrix), and {\bf b} is the last column of {\bf M} whose upper $k-1$ entries are equal to 1 and lower $k-1$ entries are equal to 0. An example of such matrix {\bf M} is given in Figure 1 (right)
where the column {\bf b} is marked in boldface.
$$
\left[\begin{array}{ccccc}
1&1&1&0&0\\
0&1&1&1&0\\
0&0&1&1&1\\
1&0&0&1&1\\
1&1&0&0&1\\
1&1&1&0&0\\
0&1&1&1&0\\
0&0&1&1&1\\
1&0&0&1&1\\
1&1&0&0&1\\
\end{array}\right]
\qquad\qquad
\left[\begin{array}{cccccc}
1&1&1&0&0&{\bf 0}\\
0&1&1&1&0&{\bf 0}\\
0&0&1&1&1&{\bf 0}\\
1&0&0&1&1&{\bf 0}\\
1&1&0&0&1&{\bf 0}\\
1&1&1&0&0&{\bf 1}\\
0&1&1&1&0&{\bf 1}\\
0&0&1&1&1&{\bf 1}\\
1&0&0&1&1&{\bf 1}\\
1&1&0&0&1&{\bf 1}\\
\end{array}\right]
$$
%
\centerline{\footnotesize Fig. 1. Left: $3$-banded matrix {\bf M} ($k=5$, $l=3$).
Right: augmented $3$-banded matrix {\bf M} ($k=6$, $l=3$).}


\begin{theorem}\label{th2} The $l$-banded matrix ${\bf M}$ satisfies the assignment property. 
\end{theorem}
{\sc Proof.} We use Lemma~\ref{lem1}. Fix some $m\in\{1,\dots, k\}$ and a sequence of different indices $i_1,\dots, i_m$. For brevity we will write  ${\bf A}(i_1,\dots, i_m)={\bf A}$. Denote by $r$ the number of non-null columns of ${\bf A}$. We denote by  ${\bf M}_{\bf u}$  the upper  $k\times k$ submatrix of {\bf M}  and by ${\bf M}_{\bf l}$  the lower $k\times k$ submatrix of {\bf M} . The matrices  ${\bf M}_{\bf u}$  and  ${\bf M}_{\bf l}$  are identical. 
Let $i$ be the number of rows of ${\bf A}$ taken from the upper submatrix  ${\bf M}_{\bf u}$ and $j=m-i$ the number of rows of ${\bf A}$ taken from the lower submatrix  ${\bf M}_{\bf l}$. Assume without loss of generality that $j\le i$. Then 
\begin{equation}\label{1}
j\le l-1.
\end{equation}
Indeed, since $i+j=m\le k$ we have $2j\le k$ and, as $k$ is odd, $2j \leq k-1=2(l-1)$.

We now show the following fact:
\begin{equation}\label{2}
\text{\it the $r$ null columns of {\bf A} form necessarily a group of $r$ consecutive columns} 
\end{equation}
(the word "consecutive" is 
understood in a cyclic way, which means that the last column of {\bf A} is the left neighbor of its first column).  Indeed, assume that (\ref{2}) is not true, that is there exist two null columns of {\bf A} separated by non-null columns (in a cyclic way). Denote these columns by ${\bf A}_1$ and ${\bf A}_2$. Recall that rows of {\bf A} are also rows of {\bf M}.  By the definition of {\bf M}, every two zeros in any row of {\bf M} are separated by at least $l$ ones. This implies that there exists at least one row of {\bf A} containing $l$ ones in the left direction between ${\bf A}_1$ and ${\bf A}_2$, and at least one row of {\bf A} containing $l$ ones in the right direction between ${\bf A}_1$ and ${\bf A}_2$ (we consider {\bf A} as a cyclic matrix as explained above). These two properties are only compatible if the total number of columns of {\bf A} is at least $2l+1$.
Since {\bf A} has $k$ columns and $2l+1>k$, we come to a contradiction that proves (\ref{2}).

Fix now arbitrary $r$ consecutive columns of {\bf A} and assume that these are null columns. Then it is immediately clear that some rows of the upper submatrix ${\bf M}_{\bf u}$ cannot be part of  {\bf A} since they contain ones in the positions corresponding  to these columns. Let us call such rows $r$-forbidden rows.
It is easy to deduce from the definition of  ${\bf M}_{\bf u}$ that the number of $r$-forbidden rows is $r+l-1$, whatever is the position of $r$ consecutive null columns. Therefore, the total number $i$ of rows of ${\bf A}$ taken from the upper submatrix  ${\bf M}_{\bf u}$ satisfies $i\le k-(r+l-1)$. Combining this inequality with (\ref{1}) we get $m=i+j\le k-r$. Thus, the condition of Lemma~\ref{lem1} is satisfied, which proves the theorem.

The next result is related to the second construction.

\begin{theorem}\label{th3} 
The augmented $l$-banded matrix ${\bf M}$ satisfies the assignment property. 
\end{theorem}
{\sc Proof.}  We use again Lemma~\ref{lem1}. We have to show that the number of non-null columns of ${\bf A}(i_1,\dots, i_m)$ is at least $m$ for any 
$m\le k$. In view of Theorem~\ref{th2}, this condition is satisfied for $m\le k-1$. Indeed, $k'=k-1$ is even, $k'\ge 3$, and the matrix ${\bf M}'$ composed of the first $k-1$ columns of ${\bf M}$ is an $l$-banded matrix with dimensions $n\times k'$. Therefore, by Theorem~\ref{th2}, for $m\le k'$ there exist at least $m$ non-null columns among the first $k'$ columns of ${\bf A}(i_1,\dots, i_m)$.
It remains to consider the case $m=k$. Then, by the above argument, there exist already $k-1$ non-null columns of ${\bf A}(i_1,\dots, i_k)$ among its first $k-1$ columns -- we can take the non-null columns of ${\bf A}(i_1,\dots, i_{k-1})$ augmented by one element 0 or 1. Finally, note that the last column of ${\bf A}(i_1,\dots, i_k)$ is always non-null since among any chosen $k$ rows of {\bf M} there should be at least one from the lower submatrix, i.e., a row with $k$th entry equal to 1. 

\vspace{2mm}

{\sc Remark.}
We note that if we remove any $w$ rows from the proposed matrices  {\bf M}, we obtain $(n-w)\times k$ binary matrices  that still has the assignment property provided $n-w\ge k$. This means that from the first construction we can obtain matrices with the assignment property for any $n$ such that  $k\le n\le 2k$ for odd $k\ge 3$. The second construction extends in this way to any $n$ such that $k\le n\le 2(k-1)$ for even $k\ge 4$. By extension, we call these $(n-w)\times k$ matrices $l$-banded or augmented $l$-banded as well.


\section{Algorithmic realization for $l$-banded matrix}

Theorems~\ref{th2} and \ref{th3} provide the existence results but do not exhibit concrete algorithms of Walsh code assignment. Note however that, in view of the reduction to Hall's theorem, the algorithmic realization can be readily done via the algorithms of finding different representatives that are available in the literature (cf., for example, \cite{hall}). What is more, for the $l$-banded matrix {\bf M}, there exists a very simple algorithm of Walsh code assignment
that we describe in this section.

We choose arbitrary $k$ rows of the $l$-banded matrix {\bf M}. They form a $k\times k$ matrix {\bf K}. The algorithm provides a permutation of the rows of {\bf K} transforming it into a matrix with all diagonal entries 1.  To define the algorithm we will need some notation. Namely, we will attribute some labels to the rows of {\bf M}. Recall first that, by construction,  the rows $j$ and $j+k$ of $l$-banded matrix {\bf M}  are identical; the row $j+k$ will be called the duplicate of row $j$ and vice versa.
The row $j$ belongs to the upper submatrix ${\bf M}_{\bf u}$ while the row $j+k$ belongs to the lower submatrix ${\bf M}_{\bf l}$. 
\begin{itemize}
\item If neither of the rows $j$ and $j+k$ of {\bf M} is chosen for {\bf K}, these rows are called {\it void rows} (or {\it V-rows}) and they are marked by label $V$.
\item If only one of the rows $j$ and $j+k$ of {\bf M} is chosen for {\bf K}, this row is called a {\it single row} (or {\it S-row}) and it is 
marked by label $S$.
\item If both rows $j$ and $j+k$ of {\bf M} are chosen for {\bf K}, these rows are called {\it double rows} (or {\it D-rows}) and they are marked by label $D$.
\end{itemize}
Note that the number of $D$-rows in the upper submatrix is equal to the number of $V$-rows. 
It will be convenient to consider at the beginning the matrix {\bf K} within the matrix {\bf M}, with the rows of {\bf M} marked as defined above. 

If all the rows of {\bf K} are $S$-rows, then there is no problem and the desired permutation is straightforward:  we just move the rows of {\bf K} chosen from the lower submatrix ${\bf M}_{\bf l}$ to the identical positions in the upper submatrix ${\bf M}_{\bf u}$. As a result, the upper submatrix will become a $k\times k$ matrix with all diagonal entries 1 and therefore it will be the desired output matrix. In general, when there are also $V$ and $D$ rows, we will need to 
perform additional permutations of the marked rows of {\bf M} but we will still define the output of our algorithm as a $k\times k$ matrix obtained in place of ${\bf M}_{\bf u}$ at the last step of the algorithm. 

The definition of the algorithm is as follows. First, if there are $S$-rows of {\bf K} in the lower submatrix ${\bf M}_{\bf l}$, move them to the identical positions in the upper submatrix ${\bf M}_{\bf u}$. As a result, all the rows of the upper submatrix are now marked by labels $V,S$ or $D$. Note that the number of $D$-rows in the upper submatrix is equal to the number of $V$-rows. For any $D$-row, consider the closest $V$-row from below (in a cyclic way, so that the uppermost row of the upper submatrix is viewed as the first below its last row). Form a couple of these two $D$ and $V$-rows; couple in the same way all the other $D$ and $V$-rows. To each ($D$,$V$) couple we associate a cluster of consecutive rows of the upper submatrix composed of these two rows and all the $S$-rows between them. To finish the algorithm, we process each cluster in the following way (working with the cyclic upper submatrix as above).
\begin{itemize}
\item[(i)] If the $V$-row of the couple stands next below its $D$-row (no $S$-rows in between), replace this $V$-row by the duplicate of the $D$-row.
\item[(ii)]  If there are $S$-rows between the $D$ and $V$-rows of the couple (further called $S$-rows of the cluster), put the duplicate of the $D$-row next below the $D$-row of the couple and shift all the $S$-rows of the cluster one row down. 
\end{itemize}
The output of the algorithm is the upper submatrix obtained after performing these operations.
It is easy to see that all its diagonal entries are equal to 1.  Indeed, the $S$-rows outside the clusters are not modified by (i) and (ii) and they have ones on the diagonal positions.   
Next, operations (i) and (ii) eliminate the $V$ rows, move in the duplicates of $D$-rows, so that the resulting upper submatrix is a permutation of the rows of {\bf K}. Finally, operations (i) and (ii) yield diagonal entries 1 in each cluster. Indeed, in $S$ and $D$ rows, the second entry right from the diagonal is 1 since for $k\ge3$ we have $l=(k+1)/2\ge2$.

Note that the complexity (i.e., is the number of operations) of this algorithm is not high. 
We need to shift some of the rows of a chosen $k\times k$ matrix {\bf K} up to the point when we obtain a matrix with all diagonal entries 1; each row is shifted at most once. This means that the complexity of the algorithm is at most $k$.


\section{Optimality of the assignment matrices}

Apart from the matrices {\bf M} proposed in this paper, there exist many other matrices satisfying the assignment property. For example, one can take the $n\times k$  matrix with all entries~1.  However, this matrix contains too many ones, so that the initial assignment table $S$ becomes too large; we need to have $l=k$. On the other hand, for our constructions above, $l$ is approximately $k/2$.  A natural question is whether one can find matrices {\bf M} with even smaller $l$. We will show that this is impossible and thus our suggestions are optimal (the most parsimonious) with respect to the criterion defined as the number of entries 1 in the matrix. 

The following theorem gives a necessary condition for the assignment property.
\begin{theorem}\label{th4} Let ${\bf M}$ be any binary $n\times k$ matrix with $n\ge k$, and let $N$ denote the total number of entries 1 in ${\bf M}$.
Then the condition
\begin{equation}\label{4}
N \ge k(n-k+1)
\end{equation}
is necessary for the assignment property of ${\bf M}$. 
\end{theorem}
{\sc Proof.} Assume that ${\bf M}$ satisfies the assignment property. Then the number of entries 0 in each column of ${\bf M}$ is less than or equal to $k-1$.
Indeed, assume that there exists a column of ${\bf M}$ with $k$ or more entries 0. Compose a $k\times k$ matrix ${\bf K}$ from the corresponding rows of ${\bf M}$. Clearly, by permutation of the rows of ${\bf K}$ we cannot obtain a matrix with all diagonal entries 1. Thus, the assignment property does not hold and we come to a contradiction.
As a consequence, the number of entries 1 in each column of ${\bf M}$ is greater than or equal to $n-k+1$, which implies the result of the theorem.

\vspace{1mm}

As a consequence of (\ref{4}) we get
\begin{equation}\label{5}
n l_{\max} \ge k(n-k+1)
\end{equation}
where $l_{\max}$ is the maximal number of ones in a row of ${\bf M}$.

A binary $n\times k$ matrix {\bf M} is called {\it optimal} if the lower bound (\ref{4}) is attained, that is $N=k(n-k+1)$. It is easy to see that the $l$-banded matrix {\bf M} is optimal in the sense of this definition. Indeed, for the $l$-banded matrix we have $n=2k$, $N=2kl=k(k+1)$.  This shows that the lower bound (\ref{4}) is tight for even $k\ge3$ and $n=2k$. The question whether it is tight in general remains open. 
Note also that for the $l$-banded matrix the number of ones in all the rows is the same and equals $l_{\max}=l$. Thus, in view of (\ref{5}), the $l$-banded matrix has the minimal number $l$ of Walsh codes assigned to each user for even $k\ge3$ and $n=2k$. 

For the augmented $l$-banded matrix we have $n=2(k-1)$ and $N=2(k-1)l +(k-1)=(k-1)(k+1)$ while the lower bound (\ref{4}) is $k(k-1)$. So, the lower bound is not exactly attained but the ratio of the upper and lower bounds tends to 1 as $k$ becomes large.


\section{Conclusion}

In this paper, we considered the problem of orthogonal variable spreading Walsh code assignments. The aim was to present such assignments that can avoid complicated signaling from BS to users or blind rate and code detection if there is a large number of possible codes out of which BS has to choose a code for transmission.  The assignments considered here use a partition of all users into several pools. Each pool can use its own codes that are different for different pools and are taken from the general set of available codes. Each user has only a few codes assigned to him.  

We stated this as a combinatorial problem expressed in terms of a binary matrix  {\bf M}.  A solution of the problem is given by a matrix {\bf M} having a specific property that we have named the assignment property. Two constructions of  {\bf M} satisfying this property  are presented
under different conditions on the number $n$ of users in a pool and on the number $k$ of  Walsh codes in the pool.
 The first of the proposed matrices {\bf M}  is optimal in the sense that it has the minimal number $l$ of Walsh codes assigned to each user for certain $n$  and  $k$. The second matrix is near optimal in the same sense. The optimality follows from the proved
 necessary condition of the existence of  {\bf M}  with the assignment property. 
 
	Additionally, we proposed a simple algorithm of assignment for the first matrix. Its complexity is not high. The complexity is the number of operations needed to find a suitable code channel assignment for the $k$ chosen MSs. It was shown that to determine the code channel assignment, the BS should shift some of the rows of a chosen $k\times k$ matrix {\bf K} up to the point when they obtain a matrix with all diagonal entries 1; each row is shifted at most once. This means that the complexity of the algorithm is at most $k$.
	
Our constructions of {\bf M}  with the assignment property are not unique. For example, if {\bf M}  is a matrix having the assignment property, then any matrix  obtained by a row and/or column permutation of {\bf M} has the assignment property as well.

\vspace{4mm}

{\bf Acknowledgement.} We would like to thank L.A.Bassalygo and the referee for their comments leading to a substantial improvement of the paper. The first author expresses gratitude to Edward Tiedemann and Peter Gall for helpful discussions.


\begin{thebibliography}{99}

\bibitem{vit} A.J.Viterbi. {\it CDMA: Principles of spread spectrum communication.} Addison-Weseley, New York, 1995.
\bibitem{2} F.Adachi, M.Sawahashi, K.Okawa. Tree-structured generation of orthogonal spreading with different lengths for forward link of DS-CDMA mobile radio. {\it Electron. Lett.}, vol.33, n.1, 27-28, 1997. 
\bibitem{3} E.H.Dinan, B.Jabbari. Sreading codes for direct sequence CDMA and Wideband CDMA Cellular Networks. {\it IEEE Communications Magazine} 48-59, September 1998.
\bibitem{4} T.Minn, K-Y.Siu. Dynamic assignment of orthogonal variable-spreading-factor codes in W-CDMA. {\it IEEE Journal on Selected Areas in Communications}, vol.18, n.8, 1429-1440, 2000.
\bibitem{5} Y.-C.Tseng, C.-M.Chao, S.-L.Wu. Code placement and replacement strategies for wideband CDMA OVSF code tree management. {\it IEEE Global Telecommunications Conference (GLOBECOM '01)}, vol.1, 562-566, 2001.
\bibitem{6} M.Dell'Amico, M.L.Merani, F.Maffioli. Efficient algorithms for the assignment of OVSF codes in wideband CDMA. {\it IEEE International Conference on Communications (ICC 20020}, vol.5, 3055-3060, 2002.
\bibitem{7} S.Tsai, F.Khaleghi, S.-J. Oh, V.Vanghi. Allocation of Walsh codes and quasi-orthogonal functions in cdma2000 forward link. {\it IEEE Vehicular Technology Conference (VTC 2001)}, vol.2, 747-751, 2001. 
\bibitem{hall} M. Hall. {\it Combinatorial Theory.} New York, Wiley, 1998.




\end{thebibliography}
\end{document}